\documentclass[11pt]{article}
\usepackage[dvips]{color}
\usepackage{mathtext}
\usepackage[koi8-r]{inputenc}
\usepackage{graphicx}
\usepackage[figuresright]{rotating}
\usepackage{amssymb,eucal}
\usepackage{cite}

\newcommand{\beq}{\begin{equation}}
\newcommand{\eeq}{\end{equation}}
\newcommand{\beqn}{\begin{eqnarray}}
\newcommand{\eeqn}{\end{eqnarray}}

\date{}
\begin{document}
\title{Scintillator counters with WLS fiber/MPPC readout for the side muon range 
detector (SMRD) \\
  of the T2K experiment}

\author{A.~Izmaylov$^a$\footnote{Corresponding author.
{\it Email address:} izmaylov@inr.ru}, S.~Aoki$^b$, J.~Blocki$^c$,
J.~Brinson$^d$, A.~Dabrowska$^c$,\\  
I.~Danko$^e$,  M.~Dziewiecki$^f$, B.~Ellison$^d$, L.~Golyshkin$^a$, R.~Gould$^d$, \\ 
T.~Hara$^b$, B.~Hartfiel$^d$, 
J.~Holeczek$^j$,   M.~Khabibullin$^a$, A.~Khotjantsev$^a$, \\ 
D.~Kielczewska$^i$, J.~Kisiel$^j$,  T.~Kozlowski$^h$, Yu.~Kudenko$^a$, R.~Kurjata$^f$,\\
T.~Kutter$^d$, J.~Lagoda$^h$,  J.~Liu$^d$, J.~Marzec$^f$, W.~Metcalf$^d$, P.~Mijakowski$^h$, \\ O.~Mineev$^a$,  Yu.~Musienko$^a$, D.~Naples$^e$, M.~Nauman$^d$, D.~Northacker$^e$, \\ J.~Nowak$^d$, V.~Paolone$^e$, M.~Posiadala$^i$, P.~Przewlocki$^h$, J.~Reid$^d$, \\ 
E.~Rondio$^h$, A.~Shaykhiev$^a$, M.~Sienkiewicz$^c$, D.~Smith$^d$, J.~Sobczyk$^k$,  \\ M.~Stodulski$^c$, A.~Straczek$^c$, R.~Sulej$^f$, A.~Suzuki$^b$, J.~Swierblewski$^c$, \\ T.~Szeglowski$^j$, M.~Szeptycka$^h$, T.~Wachala$^c$, D.~Warner$^g$, N.~Yershov$^a$, \\ T.~Yano$^b$, A.~Zalewska~$^c$, K.~Zaremba$^f$, M.~Ziembicki$^f$ \\
 {} \\
$^a${\it Institute for Nuclear Research RAS, Moscow 117312, Russia} \\ 
$^b${\it Kobe University, Kobe, Hyogo 657-8501 , Japan} \\
$^c${\it H. Niewodniczanski Institute of Nuclear Physics PAN,} \\
{\it  Krakow 31-342, Poland } \\
$^d${\it Louisiana State University, Baton Rouge, LA 70803, USA} \\
$^e${\it University of Pittsburgh, Pittsburgh, PA 15260, USA} \\
$^f${\it Institute of Radioelectronics, Warsaw University of Technology,} \\
{\it Warsaw 00-665, Poland} \\
$^g${\it Colorado State University, Fort Collins, CO 80523, USA} \\
$^h${\it A. Soltan Institute of Nuclear Studies, Warsaw 00-681, Poland} \\
$^i${\it Institute of Experimental Physics, University of Warsaw,} \\
{\it Warsaw 00-681, Poland} \\
$^j${\it {Institute of Physics, University of Silesia, Katowice 40-007, Poland}}  \\
$^k${\it Institute of Theoretical Physics, Wroclaw University,} \\
{\it  Wroclaw 50-204, Poland}
{}\\
{}\\
{}\\
{}\\
{}\\
{}\\
{}\\}
\maketitle

%
%
%
%

\begin{abstract}
The T2K neutrino experiment at J-PARC uses a set of near detectors to 
measure the properties of an unoscillated neutrino beam and neutrino 
interaction cross-sections.  
One of the sub-detectors of the near-detector complex, the side muon range 
detector (SMRD),  is described in the paper. The detector is designed to 
help measure the neutrino energy spectrum, to identify background and to 
calibrate the other detectors. The active elements of the SMRD consist 
of 0.7~cm thick extruded scintillator slabs inserted into air gaps of 
the UA1 magnet yokes. The readout of each scintillator slab is provided 
through a single WLS fiber embedded into a serpentine shaped groove. 
Two Hamamatsu multi--pixel avalanche
photodiodes (MPPC's) are coupled to both ends of the WLS fiber. This design 
allows us to achieve a high MIP detection efficiency of greater than 99\%.
A light yield of 25-50 p.e./MIP, a time resolution of about 1 ns  
and a spatial resolution along the slab better than 10 cm were obtained for the 
SMRD counters.     

\end{abstract}

%
%

\section{Introduction}
\begin{figure}[hbt]
\begin{center}
\includegraphics[scale=0.4]{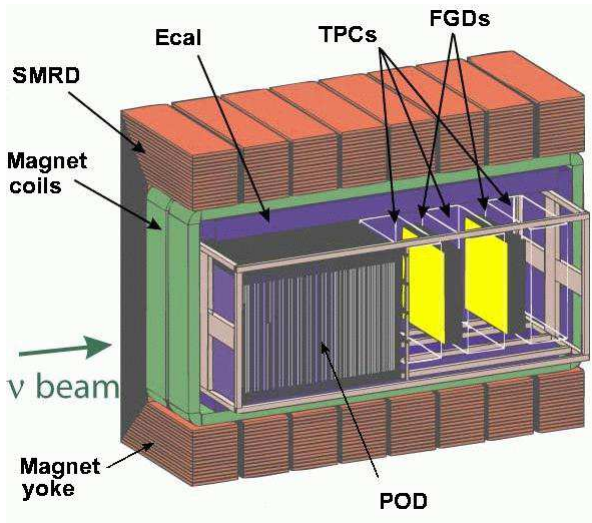}
\end{center}
\caption{
\label{fig:ndet}
Schematic view of the ND280 detector. One side has been removed for clarity.}
\end{figure}

The T2K neutrino experiment~\cite{t2k} is a second generation long-baseline 
neutrino experiment and its primary goal is to measure the unknown neutrino 
mixing parameter $\theta_{13}$. A muon neutrino beam generated by the 50 GeV 
(initially 30 GeV) proton synchrotron at	
J-PARC (Japan Proton Accelerator Research Complex) is directed towards the 
Super-Kamiokande detector located 295 km away. The complex of near 
detectors, ND280~\cite{nd280_1,nd280_2,nd280_3},  is designed to measure 
the properties of the unoscillated $\nu_{\mu}$ beam and neutrino 
interaction cross-sections. It is located 280 m  downstream from the primary  
production target and consists of two detectors: the on-axis detector 
(neutrino monitor) and the off-axis detector. The off-axis detector operates 
with a magnetic field of 0.2 T using the UA1/NOMAD CERN magnet and consists
 of a number of sub-detectors (Fig.~\ref{fig:ndet}).

 In this paper we discuss one of the ND280 sub-detectors - the side muon range detector (SMRD). The principal goals of the SMRD are (1) to measure the muon momentum and angle for charged current quasi-elastic CC-QE reactions to help  determine neutrino energy, (2) to identify backgrounds from beam neutrino interactions in magnet yokes and surrounding walls  and (3) to provide a cosmic trigger signal for  calibration of the inner detectors.  As for point (1) the detector is designed to detect lateral muons that are unseen by the inner detectors. According to MC simulation approximately 40\% of muons from CC-QE interactions and about 15\% of muons from CC non-QE reactions are expected to intersect the SMRD.

 A high muon detection efficiency, a highly hermetic layout and long term stability are the main requirements for the detector. The time and spatial resolution of the entire SMRD strongly depend on the DAQ electronics performance. In this paper we will focus mainly on the design and tests of the individual SMRD counters.

\section{SMRD design}
The UA1 magnet yoke consists of 16 C-shaped elements (8 rings). A C-element  
is segmented into 12  azimuthal sections, each of them is made of sixteen 
48 mm thick  iron plates with 17 mm air gaps between them. Some of the 
air gaps are instrumented with the  SMRD detectors. 

In order to allow for maximum flexibility during the installation process a 
modular approach was chosen: 4 (horizontal gaps) or 5 (vertical gaps) 
individual counters are assembled by means of H-profiles to form a SMRD 
module. The total number of individual counters (including spares) is 2130.

\section{SMRD individual counters}
\subsection{Design}
\begin{figure}[hbt]
\begin{center}
\includegraphics[scale=0.3]{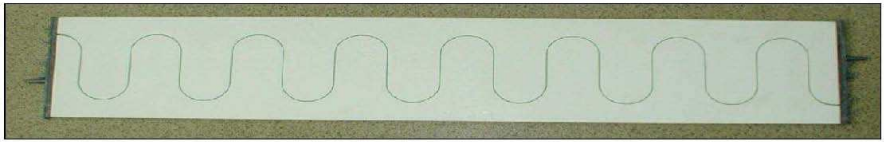}
\end{center}
\caption{
\label{fig:counter}
SMRD counter with embedded WLS fiber before wrapping into a stainless steel container.}
\end{figure}

Polystyrene based scintillator slabs with double-ended WLS fiber readout are 
used as active elements of the SMRD detector~(Fig.~\ref{fig:counter}). 
A single WLS fiber is glued into a serpentine-routed 
groove with BC600 Bicron glue~\cite{glueing}. Such a configuration allows us 
to collect scintillation light uniformly over the entire plastic surface, to 
obtain a high light yield as 
well as to minimize the number of photosensors and electronics channels  
to a pair per a counter. 

Molded end-caps are attached to the counter ends to provide a housing for 
photodetectors and temperature sensors. 
Hamamatsu 667-pixel  MPPC`s~\cite{mppc1,mppc2,mppc3} are used in the SMRD  
as photosensors. These multipixel avalanche photodiodes have a sensitive 
area of 1.3$\times$1.3 mm$^2$  and have the advantage of being insensitive to 
a magnetic field. 

To fill the magnet air gaps, scintillator  slabs of two sizes have been 
manufactured: 875$\times$167$\times$7~mm$^3$ for horizontal gaps, 
and 875$\times$175$\times$7~mm$^3$ for vertical ones.   

Extruded polystyrene scintillators have been produced  by Uniplast company 
in Vladimir, Russia. Outer surfaces of a slab are etched, thus resulting in 
formation of a white diffuse layer which acts as a reflector and  has been 
demonstrated to have an excellent performance.
The advantage of this approach is an almost ideal contact of the reflector 
with the scintillator. Some details of the extrusion technique and the 
method of etching a scintillator with a chemical agent can be found in 
Ref.~\cite{extrusion}. Extruded scintillators of this type have shown good 
light yield stability over two years~\cite{stability}.

\subsection{WLS fibers}

The Y11 (150) Kuraray WLS fibers used in the SMRD are of flexible S-type, with double-cladding and   
1 mm diameter~\cite{fibers}. Fibers are bent into a serpentine-like geometry,  consisting of 7 loops with diameter of 58 mm. The total fiber length is about 2.2 m. 

The long term stability of bent WLS fibers  was tested with two 
photosensors MRS APD`s~\cite{mrs_apd} and a blue LED as a light source. Several 3 m long fiber samples were wound into 7 turns  to reproduce the SMRD configuration. A bending diameter of about 60 mm led to an average initial drop in light transmission quality of about 5\%, which is  
in a good agreement with Kuraray data~\cite{fibers}. Light attenuation  was measured as the ratio of the light signal at the far fiber end  to the signal at the close end. The near end signal served as a reference of LED intensity.  A straight fiber of the same length was used to calibrate photosensors.
No degradation in light attenuation has been observed over a period of more than a year.    
    
\subsection{Assembly of SMRD counters}

Scintillator slabs with glued WLS fibers and attached end-caps are wrapped into an additional reflector layer of 0.1 mm thick Tyvek paper which increases a light yield by about 15\%.
Then counters are wrapped into 0.1-0.15 mm thick stainless steel foils in order to have a good protection from light and  humidity. Stainless steel has been selected to protect the counters from a possible mechanical damage during installation.

The containers are fixed with DP-490 black epoxy glue and an adhesive tape. The light isolation  of assembled counters has been checked by measuring the signal rate for light-on and light-off modes.      

\section{Individual counter performance}

The first SMRD prototype was tested   at the KEK 12-GeV synchrotron with a 
1.4~GeV/c pion 
beam~\cite{smrd_prototype}. In these tests MRS APD`s were used as 
photodetectors. A small size beam scanned all of the counter surface. Successive data analysis  demonstrated that
a light yield (sum of both ends) of more than 12 p.e./MIP results in a detection efficiency of better than 99\%.   

\begin{figure}[hbt]
\begin{center}
\includegraphics[scale=0.40]{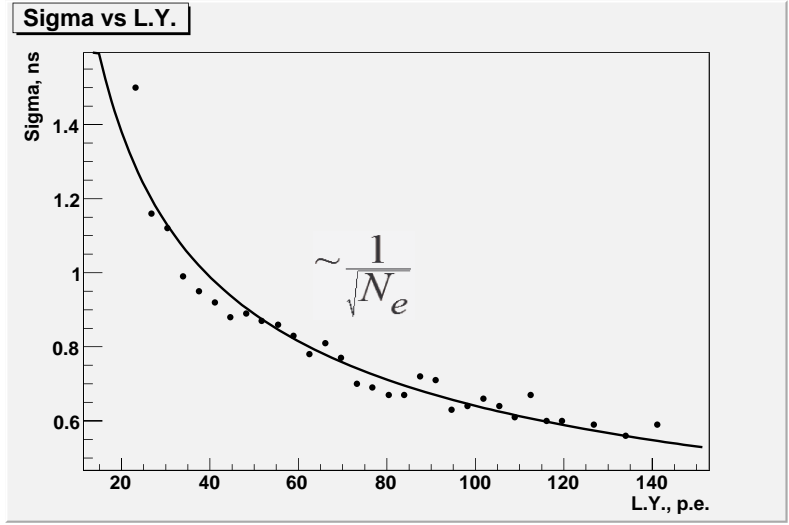}
\end{center}
\caption{
\label{fig:timing}
The time resolution after  time-amplitude correction versus the light yield for a MIP penetrating the center of the  SMRD counter.}
\end{figure}

The performance of the mass-production  SMRD counters was measured with cosmic ray muons 
and MPPC photodiodes using a small 2$\times$2 cm$^2$ trigger counter 
placed at the slab center. The mean light yield  was near  
40 p.e./MIP (sum of both ends). A high light yield of 40 p.e. corresponded 
to a timing resolution of about 0.9 ns (Fig.~\ref{fig:timing}). 
To suppress the timing spread caused by trigger counters we used 
the $(T_{left}-T_{right})/2$ combination for timing measurements. 
The corresponding spatial resolution along the slab was about 6 cm rms. 
A MIP detection efficiency of more than 99.9\% was achieved.

\section{Cosmic muon tests of SMRD counters}
\begin{figure}[hbt]
\begin{center}
\includegraphics[scale=0.55]{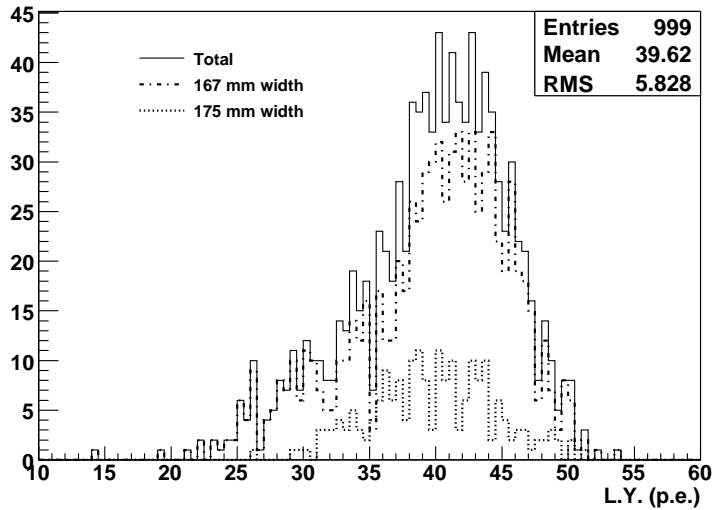}
\end{center} 
\caption{
\label{fig:ly}
The light yield distribution for about 1000 SMRD counters.}
\end{figure}

Before installation into the UA1 magnet yokes all the counters were tested with cosmic muons. The T2K experiment will collect data for almost ten years, so we have set strict limits to accept a counter for installation. The light yield requirement is more than 25 p.e./MIP for the sum of both ends and  at 20 $^o$C. The latter value corresponds to  expected  temperature inside UA1 yokes. According to our measurements the light yield ranges from 25 to 50~ p.e./MIP  (Fig.~\ref{fig:ly}).

 The asymmetry in signal size between ends was
 required to be less than 10$\%$ to guarantee no fiber damage.
 
\section{Conclusion}
The SMRD muon detector of the near-detector complex of the T2K experiment 
is described in the paper. Active elements of the SMRD are extruded 
scintillator slabs with double-ended WLS fibers and MPPC readout. 
The main feature of the detector is the usage of S-shaped grooves 
for embedding the fiber. The performance of  individual SMRD counters 
was measured with cosmic ray muons. The measured light yield value was 
25-50 p.e./MIP at 20-22 $^o$C. This light yield allows us to achieve a MIP 
detection efficiency  $> 99$\%, 
a time resolution of about 1 ns and a spatial resolution along the slab 
better than 10 cm. 

The first T2K physics run is expected in December~2009.

\section*{Acknowledgements}
 This research was supported in part by  the ``Neutrino Physics'' Program 
 of the Russian Academy of  Sciences, the RFBR (Russia)/JSPS (Japan) 
 grant \#08-02-91206 and US Department of Energy. 

The Polish SMRD groups acknowledge the support of the Ministry of Science  and Higher Education in Poland:
35/N-T2K/2007/0, PBZ/MNiSW/07/2006/36, 1 P03B 041 30 and N N202 0299 33.

%
%
%
%
%
%

\end{document}